\documentclass[a4paper]{article}
\usepackage{INTERSPEECH2019,amsmath,graphicx}
\usepackage{bm}
\usepackage{xcolor}
\usepackage{subfig}
\usepackage{lipsum}
\usepackage{hhline}
\usepackage{adjustbox}

\def\x{{\mathbf x}}

\def\R{{\mathbb{R} }}
\renewcommand{\vec}[1]{\bm{#1}}
\definecolor{cb_green}{rgb}{0.3,0.67,0.29}
\definecolor{cb_red}{rgb}{0.89,0.1,0.11}
\newcommand\todo[1]{}

\title{Multi-Frame Cross-Entropy Training for Convolutional Neural Networks in Speech Recognition}
\name{Tom Sercu$^{1~*}$, Neil Mallinar$^{2~*}$}
\address{
  $^1$IBM Research AI, IBM T. J. Watson Research Center, Yorktown Heights, New York\\
  $^2$IBM Watson, New York, New York
\thanks{$*$ Equal Contribution}
  }
\begin{document}
%
\maketitle
\begin{abstract}
We introduce Multi-Frame Cross-Entropy training (MFCE) for convolutional neural network acoustic models.
Recognizing that similar to RNNs, CNNs are in nature sequence models that take variable length inputs, we propose to take as input to the CNN a part of an utterance long enough that multiple labels are predicted at once, therefore getting cross-entropy loss signal from multiple adjacent frames.
This increases the amount of label information drastically for small marginal computational cost.
We show large WER improvements on hub5 and rt02 after training on the 2000-hour Switchboard benchmark.
\end{abstract}
\section{Introduction}
\label{sec:intro}
We will study training of Convolutional Neural Networks (CNNs) as 
acoustic models in hybrid NN/HMM speech recognition.
This paper focuses on the frame-by-frame cross-entropy training stage, where we will cheaply process additional neighboring frames to predict multiple adjacent HMM targets at the same time.

An important property of CNNs we will leverage in this work, is that they can efficiently process sequences from varying length \cite{lecun1995convolutional,lecun1998gradient}. 
This property was explored for efficient sequence training of CNN acoustic models in hybrid NN/HMM speech recognition~\cite{sercu2016advances,sercu2016dense}.
We will heavily rely on this ``dense prediction'' or ``sequence prediction'' viewpoint in this paper, which we will review in section \ref{sec:seqpred}.
Proper implementation of CNNs for sequences allows processing additional neighboring frames at minimal computational cost.

Even though end-to-end models have gained in popularity in recent years, the hybrid approach still has state of the art performance on well-respected large benchmark datasets \cite{saon2017english,xiong2016achieving}.
Additionally, hybrid models have properties which can be desirable in certain applications: principled combination with language models can help for transfering trained models to new domain specific languages, the interpretable HMM states enable easy adaptation to keyword search \cite{cui2015multilingual} where many low-confidence hypotheses need to be considered, etc.
Furthermore our work relies on CNNs as variable-length sequence classification models \cite{lecun1995convolutional},
which has been adapted to end-to-end setting \cite{sainath2015cldnn,liptchinsky2017based}.
This paper does not immediately translate to the end-to-end setting,
however the core idea of processing small chunks of the utterance to get label information of part of the utterance
has been adapted to the end-to-end setting \cite{hwang2016sequence}.

For RNN acoustic models in hybrid speech recognition \cite{graves2012supervised}, it is standard to use all labels of a window during CE training \cite{xiong2016achieving,saon2017english}.
This is natural since the RNN does not reduce utterance length and generates an output for each input frame.

The rest of this paper is organized as follows: Section \ref{sec:seqpred} reviews prior work on using CNNs for sequence prediction, Section \ref{sec:ce_st} reviews standard cross-entropy vs sequence training in hybrid speech recognition, and Section \ref{sec:multiframe} introduces our new multi-frame CE training.
Related work is in Section \ref{sec:relatedwork} and experiments showing the computational and WER gains of MFCE are presented in Section \ref{sec:experiments}.

\section{Sequence prediction with CNNs}
\label{sec:seqpred}
The traditional hybrid NN/HMM viewpoint sees CNNs as a specific type of DNNs which process a fixed context of frames and predicts the target of the center frame.
RNNs at the other hand are capable of processing variable length utterances.
In this section we review the most important papers that challenge this viewpoint and present CNNs from the perspective of variable-length sequence processing.

Standard DNN acoustic models \cite{hinton2012deep} predict the phonetic content of a single frame, taking as input the frame with some surrounding number of context frames on the left and right.
Traditional DNN models require fixed context windows and flatten the features before processing.
During (cross-entropy) training, random frames from different utterances and speakers are assembled into minibatches.

In contrast, 
TDNNs \cite{waibel1989phoneme} and CNNs \cite{lecun1995convolutional,lecun1998gradient} were originally conceived to process variable-length sequences,
wherein a series of learned filters are convolved over the input data.
In introduction of modern CNNs to large-scale speech recognition \cite{abdel2012applying,sainath2013deep} this viewpoint went somewhat lost;
CNNs were typically described similar to DNNs, taking a fixed context window and predicting the target of the center frame only.

To adapt modern CNNs to ASR, three aspects need some care.
Firstly,
before the output layer, the convolutional pathway feeds into a stack of fully connected (FC) layers.
FC layers on first sight look like a problem since in their typical implementation they
need a fixed size input vector, thus a fixed size context.
However it has been pointed out \cite{sermanet2013overfeat,sercu2016dense} that FC layers
are equivalently implemented as $1 \x 1$ convolutions, except for the first one in which the kernel size equals the output size of the convolutional pathway (e.g. acts as a flattening operation).
This avoids the use of any pooling in order to flatten the output of the convolutional stack.

Secondly,
in the domain of speech, \cite{sercu2016advances} says to avoid padding the convolutional layers in the time dimension,
as this would prevent efficient full-utterance processing at test time.
Due to this, in each convolutional layer with square kernel of size $k$ the input sequence is reduced in time by $k-1$.
This explicitly ties the number of convolutional layers to the size of the context window needed to produce a single frame prediction.
We will refer to this number of input frames for a single output prediction as the intrinsic model sequence length $l_m$.
This type of model allows for efficient sequence prediction as the input sequence can be processed with a sliding window of stride 1, so that all neighboring computation from predictions on previous frames in the sequence are reused for the current frame. 

Thirdly,
for efficient sequence prediction with CNNs, we additionally replace pooling in time with time-dilated convolutions \cite{sercu2016dense,oord2016wavenet}.
The reason to avoid pooling in time is that with $p$ pooling operations of stride $s$ in time, the number of output frames will be reduced by a factor of $s^p$.
To perform sequence training, we require the resolution of the output not be downsampled.
This is possible by requiring time-stride $s=1$ everywhere (also pooling operations),
and instead of time-pooling, we dilate the convolutional kernels by a factor of $s$ in the time direction.
Dilation factors are cumulative through the network, giving the same $s^p$ behavior.

With this we now have an efficient, fully convolutional pathway from input sequence to output sequence predictions.
An input of length $l_m$ will produce a single prediction: we name this quantity $l_m$ the intrinsic model sequence length.
Processing a longer sequence $l_i > l_m$ will return predictions for the central $l_i-l_m+1$ frames.
So to process a full utterance of length $l_u$, the input sequence is padded at beginning and end
such that the CNN takes as input $l_u+l_m-1$ frames to predict $l_u$ output labels.

\section{Cross-Entropy vs Sequence Training}
\label{sec:ce_st}
In training a hybrid speech recognition system there are two phases.
The first phase is cross-entropy (CE) training, in which input sequences of a fixed length are sampled from full utterances and randomly batched together.
Let $f^r(t) \in \R^{3 \x D}$ be the ($D=64$-dimensional logmel) features of the frame at time $t$ for utterance $r$ in the training dataset.
Let each sample in a minibatch be (with $r,t$ randomly sampled)
$\vec{x}^r(t) = [f^r(t-c), \dots, f^r(t), f^r(t+1), \dots, f^r(t+c)] \in \R^{3 \x D \x l_m}$,
where we introduced context $c$ such that $2c+1=l_m$).
For each $\vec{x}^r(t)$ the CNN will produce the softmax output $s$ for the center frame:
$s^r(t) \in \R^{S}$ 
with S being the number of states, i.e. label dimensionality.
The CE loss (in this case reduces to NLL loss) is used to optimize $s^r(t)$:
$\mathcal{F}
=\sum_{r,t}\mathcal{F}^r(t)$
with per-sample CE loss
$\mathcal{F}^r(t)
= - \log [s^r(t)]_{k=y^r(t)}$,
where the label $y^r(t)$ (coming from HMM forced alignment) provides the index $k \in [1 \dots S]$ of the correct target.

The second phase is sequence training (ST) \cite{kingsbury2009lattice}, where full utterances are processed by the network
and a sequence-level objective is optimized (e.g. MPE \cite{povey2002minimum} or bMMI \cite{povey2008boosted} in our case).
This discriminative ST objective is a much closer proxy to the actual objective Word Error Rate (WER) than the frame-level Cross-Entropy loss.

During ST with a correct CNN implementation, predictions are efficiently computed on the full sequence of frames simultaneously (dense predictions \cite{sercu2016dense}).
This greatly helps reduce the overall computational cost of training as duplicate operations are avoided by sharing most computation between neighboring frames.

\todo{Tom, optional: link with outside world like using RL (SCST, gumbel) for sequence-level objective}

\section{MFCE: Multi-Frame CE Training}
\label{sec:multiframe}
\begin{figure}
    \centering
    \includegraphics[height=1in]{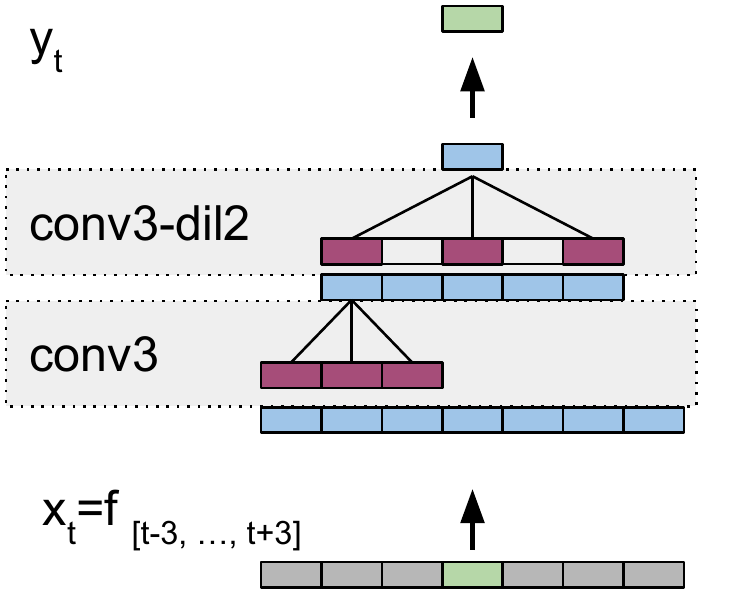}
    \caption{Simple CE training ($\delta=0$) for toy CNN with intrinsic model sequence length $l_m=7$,
    i.e. to make a single prediction the CNN has a receptive field of $l_m=7$ frames.
    }
    \label{fig:mxe_fig_1}
\end{figure}
\begin{figure}
    \centering
    \includegraphics[height=1in]{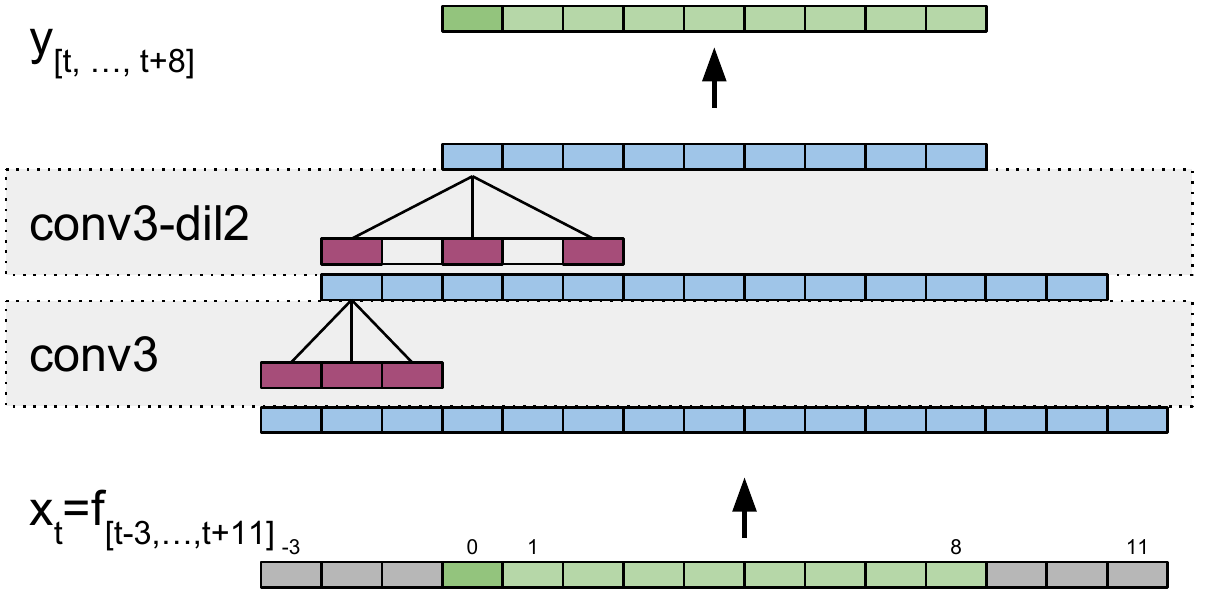}
    \caption{MFCE training for the same toy CNN with intrinsic model sequence length $l_m=7$. 
    Without any change to the CNN architecture, we add $\delta=8$ more frames to the input window,
    which makes input sequence length $l_i=l_m+\delta=15$.
    Now $(1+\delta)=9$ output states are predicted.
    It is easy to see that to add an adjacent frame, almost all computation is already done due to the convolutional structure of the network.
    Offset wrt $t$ is noted for some important input frames.
    \todo{the message would be more salient with new figure, deeper network and add smaller $\delta$}
    }
    \label{fig:mxe_fig_2}
\end{figure}

We can now introduce multi-frame cross-entropy training (or MFCE) for CNN acoustic models (Fig \ref{fig:mxe_fig_1} vs \ref{fig:mxe_fig_2}).
With multi-frame CE we process an input window of length $l_i$, which is $\delta$ longer than the intrinsic model sequence length $l_m$.
Without any modifications to the CNN architecture, this will cause the CNN to output predictions for the center $1+\delta$ frames.
Now we apply the CE loss to each of the center frames, using $1+\delta$ labels.
We introduced $\delta=l_i-l_m$ the number of additional adjacent frames.

In MFCE training, the minibatch sample $r,t$ for utterance $r$ timestep $t$ is now $\delta$ longer:
$\vec{x}^r(t) = [f^r(t-c), \dots, f^r(t+\delta+c)] \in \R^{3 \x D \x l_i}$.
The CNN predicts the sequence of $(1+\delta)$ softmax outputs $[s^r(t), \dots, s^r(t+\delta)]$.
For this sample in the minibatch, the MFCE loss becomes
$\mathcal{F}^r(t) = \frac{1}{1+\delta} \sum_{t'=t}^{t+\delta} - \log [s^r(t')]_{k=y^r(t')}$.

Using MFCE we use $(1+\delta)$ more label data per sample in the minibatch, at minimal additional computational cost.
For example, for a typical CNN which takes an input window of $l_m=53$ per prediction, we could add $\delta=8$ input frames (new input window $l_i=61$) to predict $9$ outputs at once.
The additional computational cost is roughly $\delta/l_m=15\%$ while the amount of labels used for this input increased 9-fold.

Figure \ref{fig:mxe_fig_1} and \ref{fig:mxe_fig_2} show MFCE for a small CNN.
Note that the only difference between regular CE and MFCE training is providing a longer sequence length at the CNN input; there are no architectural changes needed.
The receptive field for each output is still the same $l_m$, however the receptive field and CNN internal states are highly overlapping for adjacent output frames.

However, by training on multiple adjacent frames we also expect to have high correlation between adjacent outputs since (a) the phonetic content will be similar, and (b) the signal comes from the same speaker.
Therefore we expect the additional value of adjacent labels to be small in comparison to the labels from different samples in the batch with randomly sampled $r,t$.
The strong correlation between the labels and speakers may hurt randomization needed in a minibatch during CE training.
It is now an empirical question whether the marginal value of using the adjacent labels outweighs (a) the marginal cost of compute and (b) less random, more correlated labels.

Note that one concept requires extra care: the definition of an ``epoch'' or one pass through the data.
In this work, we define an epoch of training as a single pass over all input frames in the data. This means that, for a dataset of length $L$ frames, we can chop the input into $N = L / l_i$ samples (or input windows). This results in 
$(\delta+1)N = \frac{(\delta+1) L}{l_i} = \frac{(\delta+1) L}{l_m + \delta}$ 
labels predicted in one pass over the entire dataset.
In contrast, in the single-frame setting, with $\delta=0$, we are using less labels per epoch: only $N = \frac{L}{l_m}$.

\section{Related work}
\label{sec:relatedwork}
Somewhat closely related to our work is \cite{vanhoucke2013multiframe},
which proposes a multi-frame training strategy of DNN acoustic models.
The approach in that paper relies on using a reduced framerate, and predicting multiple states from the same input window.
Their underlying assumption is that the redundancy between adjacent input windows to predict adjacent phone states, is large enough to warrant downsampling the input windows by using stride $s \geq 2$ frames), and predict $s$ phone states from each input window.

In contrast to \cite{vanhoucke2013multiframe}, since we process the utterance fully convolutionally, we do not need to reduce the framerate by using stride $> 1$ to eliminate redundant computation.
Rather the convolutional structure allows to immediately share computation between adjacent frames without reducing the input framerate, allowing our model to make use of all frames in the utterance for prediction without incurring any computational overhead.

Another closely related work is \cite{su2013error},
where a framewise CE objective is added as regularization during ST.
As noted in Sections \ref{sec:intro} and \ref{sec:seqpred}, using multi-frame labels is standard during RNN acoustic model training. 


\section{Experiments}
\label{sec:experiments}

\todo{low priority: better to export those plots as pdf}
\todo{3 x 2 version -- larger, smaller version in comments below}
\begin{figure*}[!th]
    \centering
    \begin{tabular}{cc}
        \subfloat[Held NLL vs. Epochs]{\includegraphics[width=.38\textwidth]{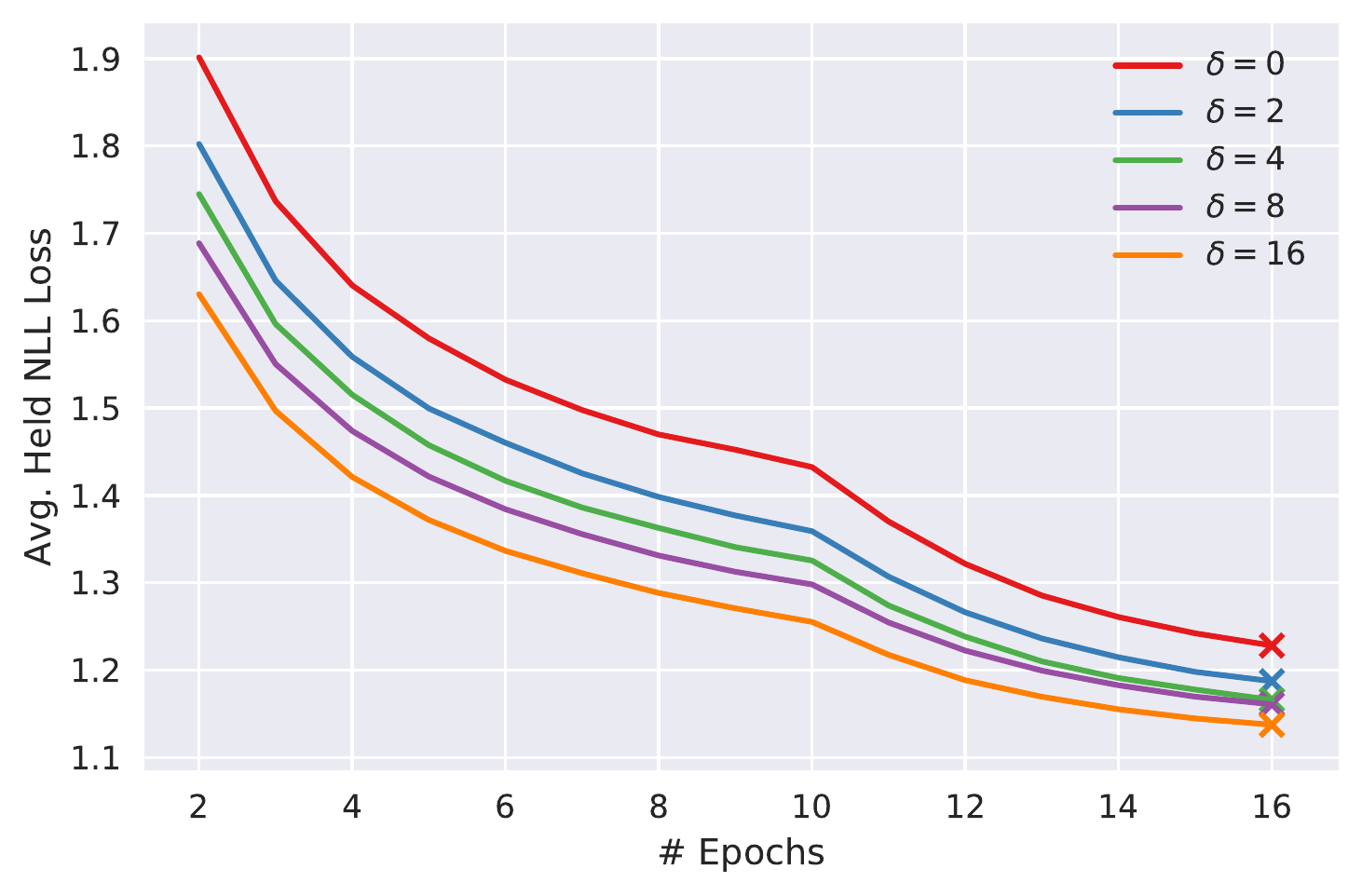} \label{fig:c_exp_nll_v_epoch}} &
        \subfloat[Held NLL vs. Time]{\includegraphics[width=.38\textwidth]{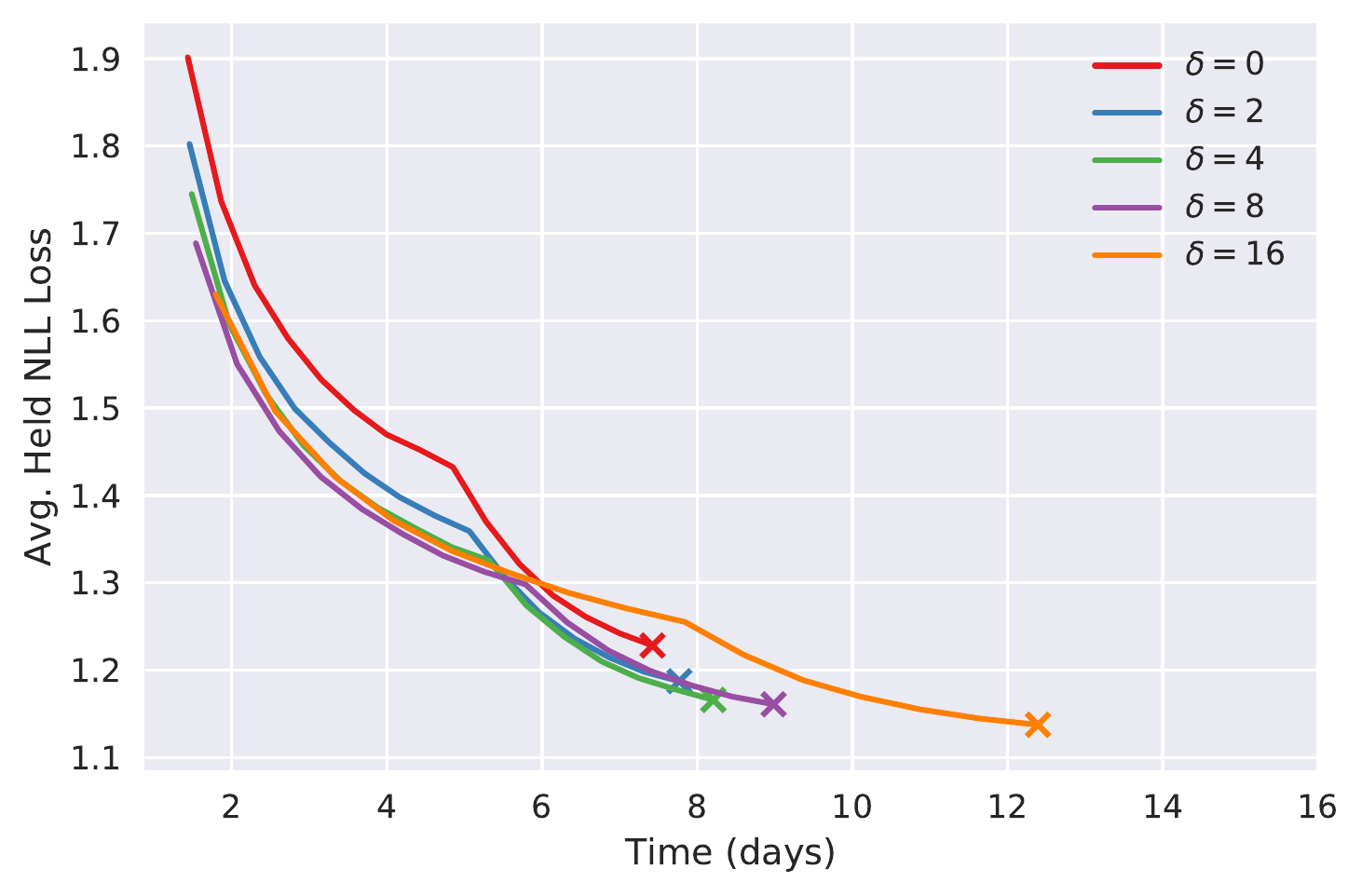} \label{fig:c_exp_nll_v_time}} \\
        \subfloat[Hub5 WER vs. Epochs]{\includegraphics[width=.38\textwidth]{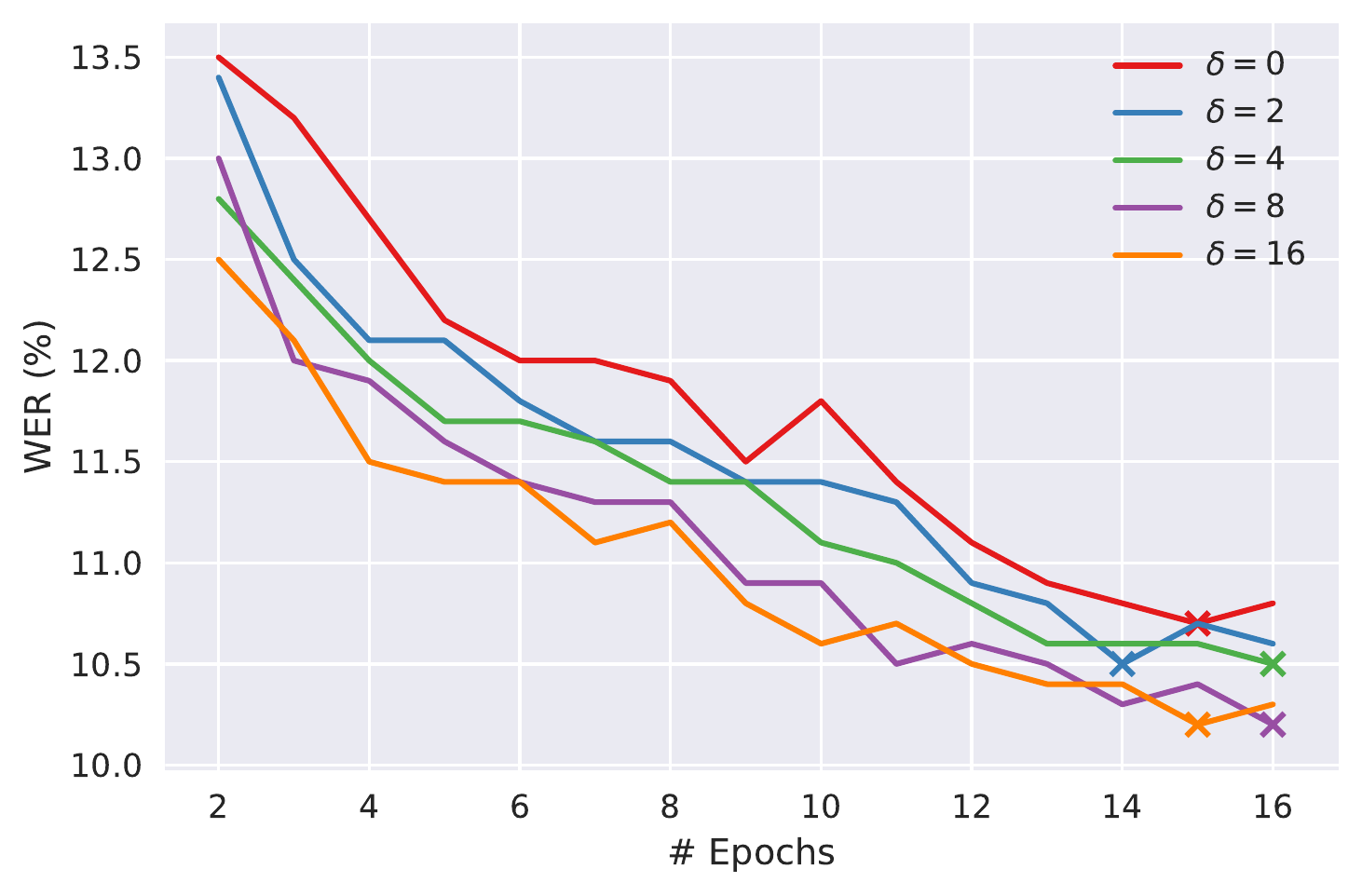} \label{fig:c_exp_hub5_wer_v_epoch}} &
        \subfloat[Hub5 WER vs. Time]{\includegraphics[width=.38\textwidth]{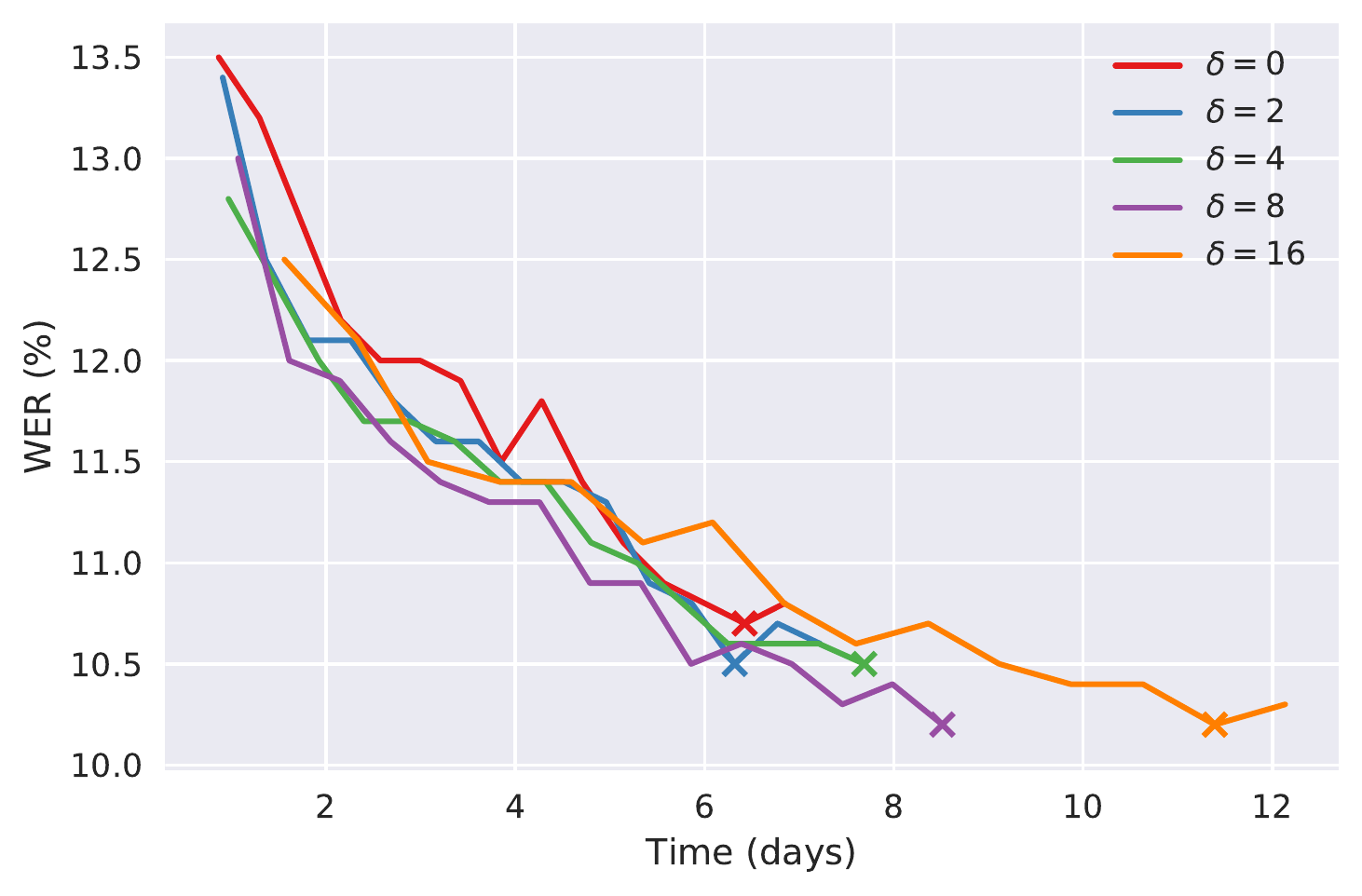} \label{fig:c_exp_hub5_wer_v_time}} \\
        \subfloat[RT02 WER vs. Epochs]{\includegraphics[width=.38\textwidth]{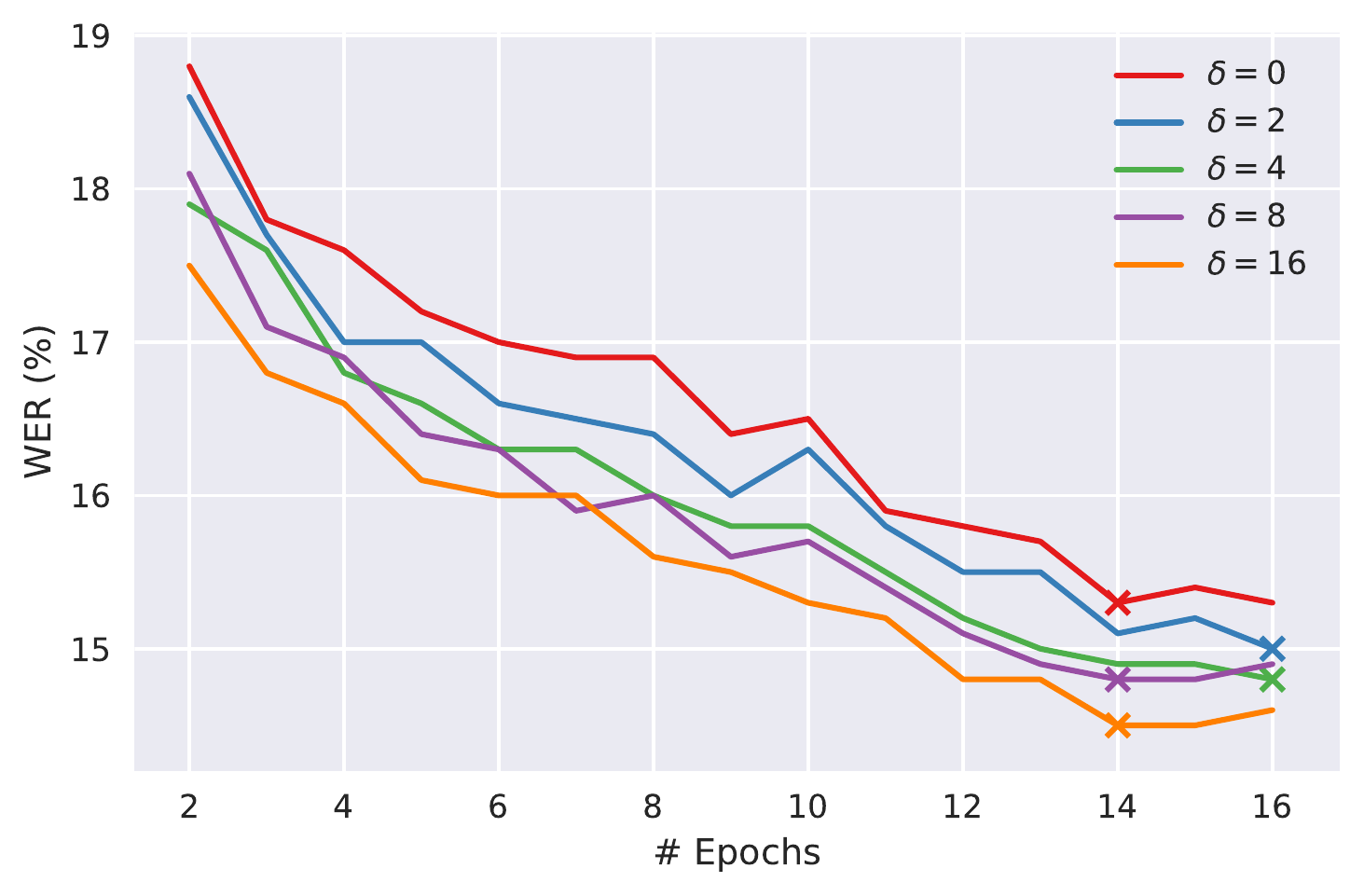} \label{fig:c_exp_rt02_wer_v_epoch}} &
        \subfloat[RT02 WER vs. Time]{\includegraphics[width=.38\textwidth]{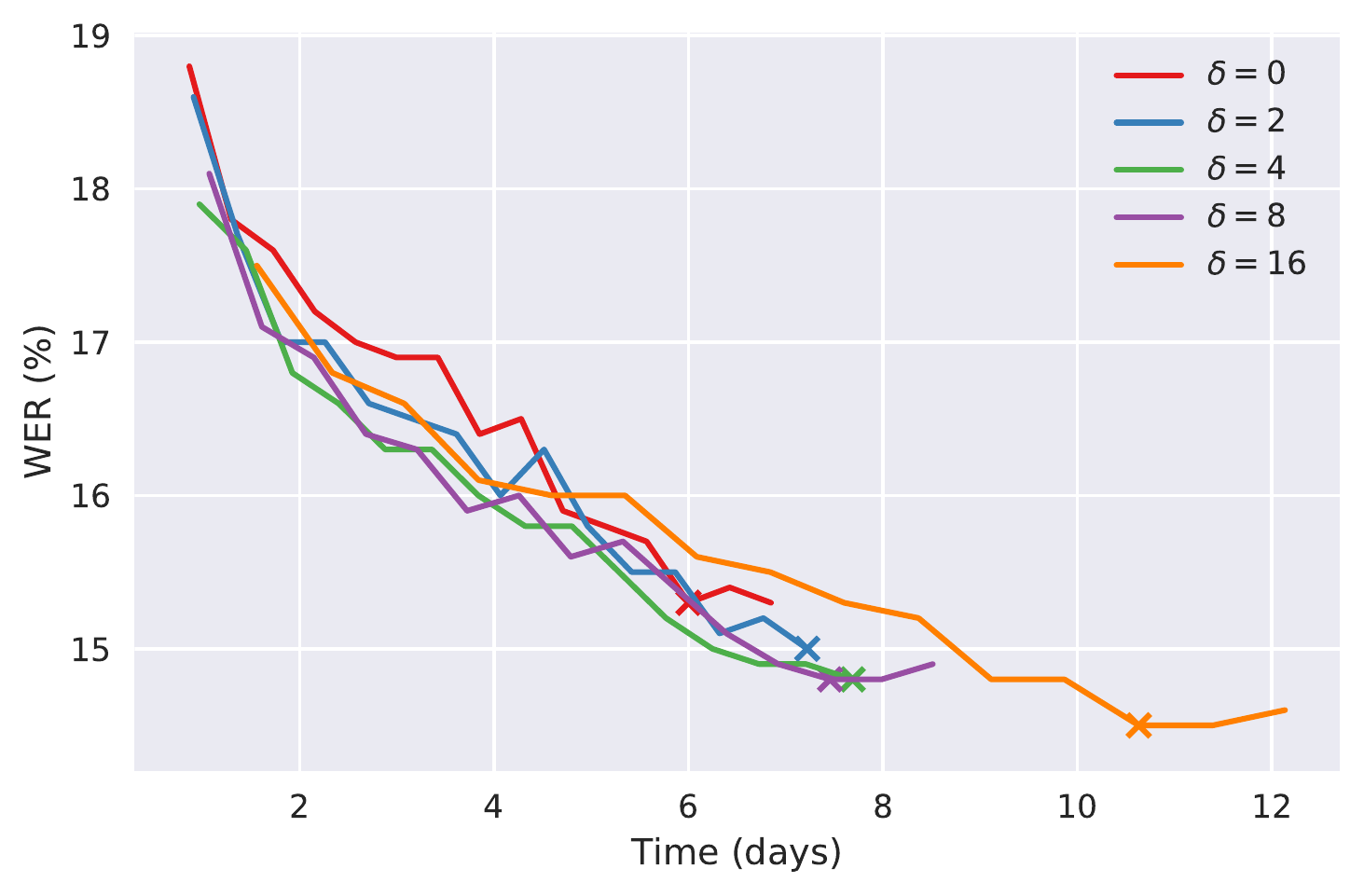} \label{fig:c_exp_rt02_wer_v_time}} \\
    \end{tabular}
    \caption{Plots of Word Error Rate (WER) and held out NLL loss during MFCE CNN training (in function of number of epochs and time in days).
    WER for two test datasets: Hub5 and rt02.
  }
    \label{fig:c_experiment}
\end{figure*}

\begin{figure*}[ht!]
    \centering
    \begin{tabular}{cc}
        \subfloat{\includegraphics[width=.38\textwidth]{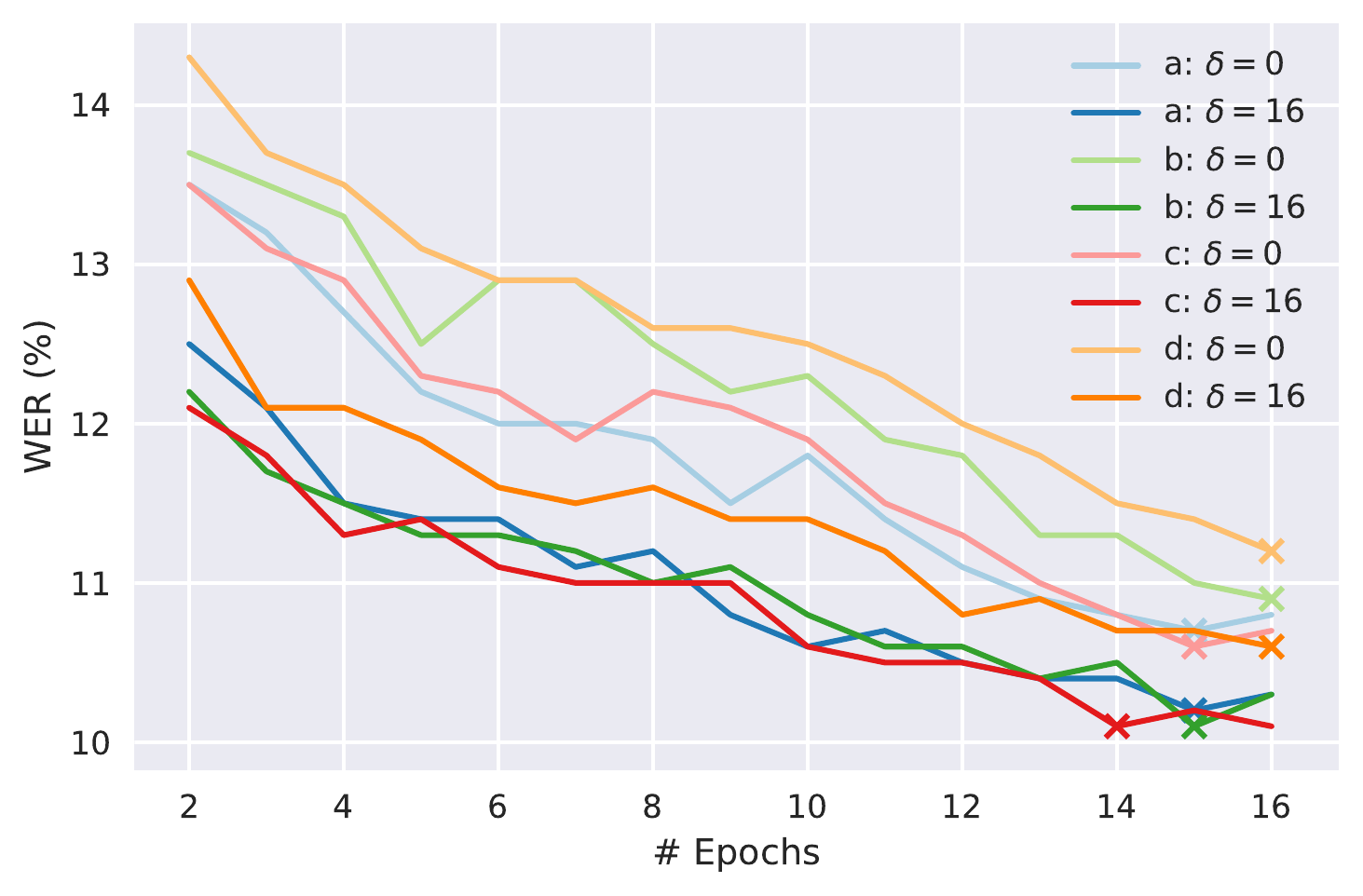}} &
        \subfloat{\includegraphics[width=.38\textwidth]{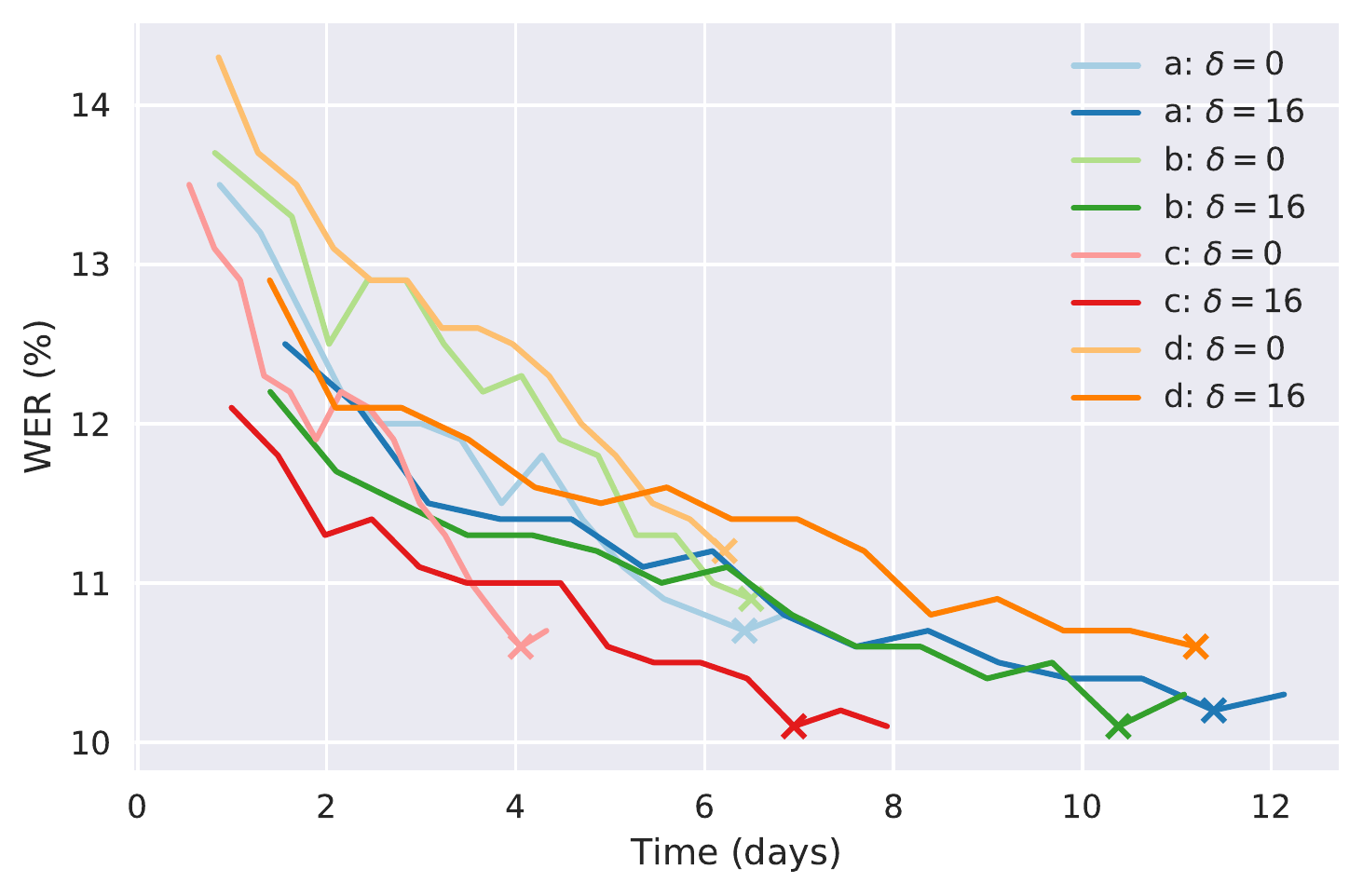}} \\
    \end{tabular}
    \caption{Comparing Hub5 WER for experimental variants that differ in per-gpu batch size, number of gpus trained on, whether or not time dilation was used in the model. Model and training details (batch is per-gpu): batch$_a=256$, n\_gpus$_a=2$; batch$_b=64$, n\_gpus$_b=2$; batch$_c=64$, n\_gpus$_c=4$; batch$_d=64$, n\_gpus$_d=4$, model $d$ has no time dilation.}
    \label{fig:experiment_variants}
\end{figure*}

\begin{table}[t]
    \centering
    \parbox[t]{4.2cm}{
        \centering
        \begin{adjustbox}{width=4.1cm}
        \small
        \begin{tabular}[t]{|l|l|l|l|}
        \hline
        Model       & NLL & Hub5 & rt02   \\ \hhline{|=|=|=|=|}
        $\delta=0$  & 1.23     & 10.7          & 15.3          \\ \hline
        $\delta=2$  & 1.19     & 10.5          & 15.0          \\ \hline
        $\delta=4$  & 1.17     & 10.5          & 14.8          \\ \hline
        $\delta=8$  & 1.16     & \textbf{10.2}          & 14.8          \\ \hline
        $\delta=16$ & \textbf{1.14}     & \textbf{10.2}          & \textbf{14.5}          \\ \hline
        \end{tabular}
        \end{adjustbox}
        \caption{Model variant $a$, varying $\delta$: best held-out NLL, Hub5 and rt02 WER (\%) selected across epochs.}
        \label{tbl:c_exp_results}
    }
    \parbox[t]{4.3cm}{
        \centering
        \begin{adjustbox}{width=4.2cm}
        \begin{tabular}[t]{|l|l|l|l|}
        \hline
        Model       & NLL & Hub5 & rt02   \\ \hhline{|=|=|=|=|}
        a: $\delta=0$  & 1.23 & 10.7      & 15.3      \\ \hline
        a: $\delta=16$ & 1.14 & 10.2      & 14.5      \\ \hline
        b: $\delta=0$  & 1.28 & 10.9      & 15.5      \\ \hline
        b: $\delta=16$ & \textbf{1.12} & \textbf{10.1}      & \textbf{14.4}      \\ \hline
        c: $\delta=0$  & 1.24 & 10.6      & 15.2      \\ \hline
        c: $\delta=16$ & 1.13 & \textbf{10.1}      & 14.5      \\ \hline
        d: $\delta=0$  & 1.57 & 11.2      & 15.9      \\ \hline
        d: $\delta=16$ & 1.49 & 10.6      & 15.3      \\ \hline
        \end{tabular}
        \end{adjustbox}
        \caption{Model variants from Figure \ref{fig:experiment_variants}:
        best NLL/WER.}
        \label{tbl:model_variant_results}
    }
\end{table}

We show results with vgg-style CNNs \cite{sercu2015very} (with time dilation \cite{sercu2016dense}) on $D=64$-dimensional logmel features with $\Delta, \Delta\Delta$ and output target space $S=32k$ tied CD states.
All models are trained in PyTorch using SGD and Nesterov acceleration with momentum 0.99, learning rate 0.01, and $L_2$ weight decay $1e^{-6}$.
We train for 16 epochs and clip gradients to $L_2$ norm 10.
From epoch 10, we begin annealing the learning rate by a factor of $\sqrt{0.5}$ per epoch.
Our model has an initial $5\x$ convolution followed by a stack of 12 $3\x3$ convolutions with 3 conv layers
having output feature maps 64, 128, 256, 512 respectively.
This architecture further follows \cite{sercu2016dense} but with as FC layers ($1\x1$ convs) only a bottleneck (512 feature maps) and output layer (32k feature maps).
By default models are trained on 2 gpus with batch size $2 \x 256$,
but we present result with variation in: per-gpu batch size, total number of gpus, and model without time dilation, in Figure \ref{fig:experiment_variants} and Table \ref{tbl:model_variant_results}.


Figure \ref{fig:c_experiment} shows our main experimental result.
In the left column we compare heldout NLL, and WER on Hub5 and rt02 against epochs; in the right column we compare against walltime (days).
Each line is the same (model $a$ in Fig \ref{fig:experiment_variants}) with per-gpu batch size 256, trained on two gpus, and $l_m=53$, differing in $\delta=0, 2, 4, 8, 16$.

In Figure \ref{fig:c_exp_nll_v_epoch}, we see that heldout NLL is consistently improved as $\delta$ increases. 
This tracks with WER improvement for both Hub5 and rt02, as seen in Figures \ref{fig:c_exp_hub5_wer_v_epoch} and \ref{fig:c_exp_rt02_wer_v_epoch}, respectively. 
On the rt02 dataset, we see in Table \ref{tbl:c_exp_results} an absolute best WER of $14.5\%$ for $\delta=16$, a $0.8\%$ gain over the usual single-frame CE trained model and $0.3\%$ over $\delta=8$.



In the right column of Figure \ref{fig:c_experiment} we show the results in terms of training walltime (on 2 NVIDIA K80 GPU's).
From Figure \ref{fig:c_exp_nll_v_time}, we see that the NLL loss in terms of training time shows a trade-off: increasing $\delta$ increases compute time (but not proportional to $\delta$!).
After 8 days of training, for example, the $\delta=4$ setup has best NLL, but eventually $\delta=8$ and $\delta=16$ achieve better results.




Note that
as function of walltime there is a cross-over point where the $\delta=16$ losses go from being below, to being above the others.
This can be attributed to learning rate decay starting later in terms of wall-time.
Since for a doubling of $\delta$, we (approximately) use twice as much labels,
we can probably decay the learning rate faster.
Speeding up the schedule according to $(1+\delta)/(l_m+\delta)$ would be equivalent to keeping the schedule constant, in terms of number of labels processed.
Somewhere in the middle between constant and this speed-up is probably optimal.
We leave this for future work.


Figure \ref{fig:experiment_variants} shows that our results hold across some hyper parameter variation and a simpler model architecture.

Tables \ref{tbl:c_exp_results} and \ref{tbl:model_variant_results} summarize the best WER/NLL values for each of the training runs from Figures \ref{fig:c_experiment} and \ref{fig:experiment_variants}, respectivey.

\section{Discussion}
We proposed MFCE for CNN acoustic models, and showed it gains significant WER performance for only a marginal increase in computational cost. 
In future work, we could explore full utterance cross-entropy training, which would require batching of utterances by length during CE training similar to ST training.
We also suggest that an adaptation of the learning rate schedule (decaying faster for lager $\delta$) could mitigate the WER performance vs wall-time trade-off identified here, and train with large $\delta$ MFCE not only to better performance, but also faster.
Lastly, one interesting extension is annealed MFCE in which $\delta$ is increased during training until we gradually reach full utterance lengths,
and simultaneously gradually transitioning from CE to the discriminative ST (bMMI) objective,
as opposed to switching from single-frame CE to full-utterance ST at a fixed point.



\vfill
\pagebreak

\bibliographystyle{IEEEtran}
\bibliography{refs}

\end{document}